# On Hadwiger's Number of a graph with partial information

Gabriel Istrate*

eAustria Research Institute, Bd. Vasile Pârvan 4, cam 045B,
Timişoara, RO-300223, Romania.
email: gabrielistrate@acm.org

August 1, 2018

## Abstract

We investigate the possibility of proving upper bounds on Hadwiger's number of a graph with partial information, mirroring several known upper bounds for the chromatic number. For each such bound we determine whether the corresponding bound for Hadwiger's number holds. Our results suggest that the "locality" of an inequality accounts for the existence of such an extension.

## 1 Introduction

The celebrated Hadwiger's conjecture [1] relates two important parameters of a graph, the chromatic number $\chi(G)$ and $h(G)$, the size of the largest clique minor of the graph. Specifically, it asserts that $\chi(G) \leq h(G)$ (or, equivalently, that any $l$-chromatic graph has $K_l$ as a minor). The conjecture has been proved in several special cases (e.g. [2]), and for random graphs [3]. Thus it is of interest to compare and prove various inequalities involving one of the two parameters. This is described by:

**Definition 1** *Let $\chi(G) \leq w(G)$ be an inequality, valid for some graph parameter $w(G)$ and all graphs $G$ from a certain class $\mathcal{C}$. We say that $h(G)$ interpolates in the previous inequality iff relation $h(G) \leq w(G)$ is valid for all graphs $G \in \mathcal{C}$. Of course, $h(G)$ might fail to interpolate in certain inequalities.*

*Supported by a PN II/"IDEI" grant from the Romanian CNCSIS and a grant from the Austrian Government.

**Example 1** *Coffman, Hakimi and Schmeichel [4] have proved (as a corollary) the following upper bound on the chromatic number of a connected graph with $m$ edges:*

$$\chi(G) \leq \frac{1}{2} + \frac{1}{4} \cdot \sqrt{\frac{1}{4} + 2m} \tag{1}$$

*The inequality that shows that Hadwiger's number is interpolating in inequality (1) is stated explicitly and proved in [5].*

Proving that a parameter is interpolating strengthens, of course, the corresponding inequalities for the chromatic number, provided Hadwiger's conjecture is true. Thus, in an intuitive sense, interpolating inequalities are weaker than non-interpolating ones, at least for special graphs, though it is, of course, possible that interpolating inequalities are tight.

The purpose of this note is to study the question of interpolation of Hadwiger's number for several upper bounds on the chromatic number due to Ershov and Kozhukin [6], Coffman, Hakimi and Schmeichel [4], Brooks [7], Welsh and Powell [8], and Stacho [9, 10].

## 2 Preliminaries

An useful reference for inequalities concerning the chromatic number is [11]. For Hadwiger's conjecture we refer the reader to [12]. We will use the standard notations $K_r$ and $C_r$ to denote the complete, respectively cycle graph with $r$ vertices. A graph $G$ is *1-reducible* to graph $H$ if the iterative removal of vertices of degree one from $G$ yields graph $H$.

**Definition 2** *Let $G$ be a graph and $e \in E(G)$ be an edge. The* contracted graph $G_e$ *is obtained from $G$ by identifying the endpoints of $e$ and joining the newly created vertex with all vertices that were adjacent to either endpoint. All other vertices and edges are the same as in $G$.*

**Definition 3** *Hadwiger's number $h(G)$ of connected graph $G$ is the largest $n$ such that one can obtain the complete graph $K_n$ from $G$ by a series of contractions.*

**Definition 4** *For a graph $G$ let $\Delta(G)$ be the largest degree of a vertex in $G$. Also define*

$$\Delta_2(G) = max_{u \in V(G)} max_{v \in N(u), d(v) \leq d(u)} d(v). \tag{2}$$

**Definition 5** *For a graph $G$ let $V_i$ be the set of vertices of degree $i$. Also define*

$$s = s(G) = \max_{i \geq \frac{\Delta(G)+2}{2}} |V_i|. \tag{3}$$

# 3 Results

The following easy result shows that the upper bound on the chromatic number due to Ershov and Kuzhukin [6] is interpolating:

**Theorem 1** *Let $G$ be a connected graph with $n$ vertices and $m$ edges. Then*

$$h(G) \leq \lfloor \frac{3 + \sqrt{9 + 8(m-n)}}{2} \rfloor. \qquad (4)$$

On the other hand, the following result shows that many upper bounds on the chromatic number are non-interpolating:

**Theorem 2** *Hadwiger's number is* not *interpolating in any of the following inequalities for the chromatic number:*

1. $\chi(G) \leq \Delta(G) + 1$ *(Brooks [7])*

2. *If the degrees of vertices in $G$ are (in decreasing order) $d_1 \geq d_2 \geq \ldots \geq d_n$ then $\chi(G) \leq \max_{1 \leq j \leq n} \min\{j, d_j + 1\}$ (Welsh-Powell [8])*

3. $\chi(G) \leq \Delta_2(G) + 1$ *(Stacho [9])*

4. $\chi(G) \leq \lceil \frac{s}{s+1}(\Delta(G) + 2) \rceil$ *(Stacho [10]),*

*even when restricted to the class of 3-colorable graphs.*

In [4] the following inequality that improves the Ershov-Kozhukhin bound on the chromatic number was proved: if $G$ is a connected graph that is not 1-reducible to a clique or an odd cycle then $\chi(G) \leq \lfloor \frac{3+\sqrt{1+8(m-n)}}{2} \rfloor$. For Hadwiger's number the improved upper bound does not quite hold. A class of counterexamples is presented in Figure 1 (a) (with $n = 8$, $m = 10$, $h(G) = 4$). It consists of four trees (edges AB, CD, EF, and GH) with connections between all trees such that no vertex is left with degree one. A similar construction works in the general case.

There is a simple technical explanation for the existence of these counterexamples: in the proof of the result from [4] Brooks' inequality is used which, as we have seen, is not interpolating. Nevertheless, we can recover the upper bound, making it interpolating by restricting the class of graphs it applies to:

**Definition 6** *An* acyclic contraction *of graph $G$ is a contraction that shrinks no cycle of $G$ to a single edge.*

For example, no edge contraction in a complete graph $K_n$, $n \geq 3$ is an acyclic contraction, since it shrinks a triangle to a single edge. On the other hand any contraction in a tree is acyclic. In particular contracting edges AB, CD, EF, and GH in the graph in Figure 1 is an acyclic contraction to complete graph $K_4$. We note that any contraction specified by a 1-reducibility is acyclic.

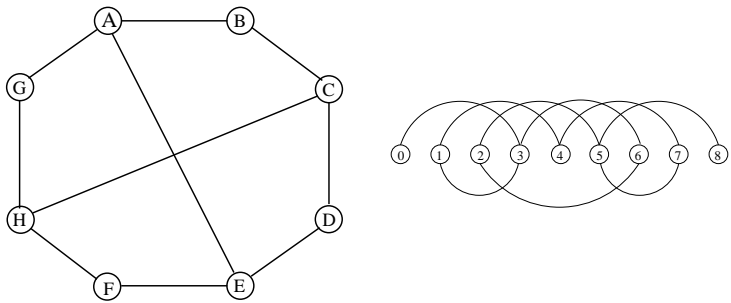

Figure 1: (a). The counterexample with $h(G) = 4$ (b). Graph $D_{3,9}$

**Observation 1** *If a graph $G$ has an acyclic contraction to an odd cycle $C_{2t+1}$ then (by additionally contracting parts of this cycle) one can obtain an acyclic contraction to $C_3 \equiv K_3$. Thus we don't really need odd cycles as a final target for acyclic contractions, avoiding acyclic contraction to a complete graph covers this case too.*

With these preliminaries we can state:

**Theorem 3** *Let $G$ be a connected graph with $n$ vertices and $m$ edges. If $G$ does not have an acyclic contraction to a complete graph $K_r$ then*

$$h(G) \leq \lfloor \frac{3 + \sqrt{1 + 8(m-n)}}{2} \rfloor.$$

Let us consider the question of obtaining interpolating upper bounds on Hadwiger's number in a graph where we know the degree sequence $d_1 \geq d_2 \geq \ldots \geq d_n$. Our results suggest an interesting difference between the interpolating and non-interpolating upper bounds: in the upper bounds in Theorem 1 and 3 the upper bounds depends on the degree sequence through quantity $m - n = \sum_i (d_i/2 - 1)$. On the other hand the non-interpolating upper bounds are "local": they essentially depend only on one of the degrees (chosen according to the specific condition of each inequality).

Of course, one can transform the sequence $\sum_i (d_i/2 - 1)$ to a "local" upper bound $n \cdot (\Delta(G)/2 - 1)$. However, this is somewhat "cheating". The question remains whether one can formalize the notion of "local" upper bounds in a way that excludes these trivial examples, and such that all tight interpolating upper bounds are *not* local (depend on $\Omega(n)$ terms in the sequence $d_i$). Such a statement is indeed true for our first result, given that interpolating inequalities in Theorem 1 and 3 are tight for all pairs $(m, n)$, $n < m < \binom{n}{2}$ (see [4]). This easily extends to the following result, which explains why some of the bounds we considered are not interpolating.

**Theorem 4** *Let $\omega = \omega_n(d_1, d_2, \ldots, d_n)$ a family of functions depending on the degrees of a graph such that:*

- *The inequality $\chi(G) \leq \omega_n(d_1, \ldots, d_n)$ is valid for all graphs $G$ in a certain class $\mathcal{C}$ with n vertices and vertex degree sequence $d_1 \geq d_2 \geq \ldots \geq d_n$.*

- *Hadwiger's number is interpolating for the above inequality.*

- *Class $\mathcal{C}$ contains graphs $D_{r,n}$ from Theorem 2*

Then $sup_{\{G \in \mathcal{C}; d_1, \ldots, d_n \leq 3\}} \omega_n(d_1, d_2, \ldots, d_n) = n^{\Omega(1)}$.

## 4 Proofs

### 4.1 Proof of Theorem 1

Consider a partition of $V(G)$ into classes $V_1, V_2, \ldots V_{h(G)}$ that correspond to vertices contracted to a specific vertex of $K_{h(G)}$. It is easy to see that the following are true:

1. Each class $V_i$ induces a connected subgraph of $G$.

2. For every $i \neq j$ there exists at least one edge connecting a vertex in $V_i$ to a vertex in $V_j$. This happens because the contracted graph $K_{h(G)}$ is a complete graph.

We have

$$\begin{aligned} m &\geq \left(\sum_{i=1}^{h(G)} |E(V_i)|\right) + \binom{h(G)}{2} \geq \left(\sum_{i=1}^{h(G)} |V_i| - 1\right) + \binom{h(G)}{2} = \\ &= [n - h(G)] + \binom{h(G)}{2}. \end{aligned}$$

i.e.
$$m \geq n + \binom{h(G)}{2} - h(G). \tag{5}$$

Thus $h^2(G) - 3h(G) - 2(m - n) \leq 0$, and the result immediately follows.

### 4.2 Proof of Theorem 2

For $r \geq 3$ we construct a graph $D_{r,n}$ with $n$ nodes, described as follows (and displayed for $r = 3, n = 9$ in Figure 1(b)):

1. Nodes are labelled 0 to $n - 1$.

2. For all $1 \leq i$, $0 \leq j \leq r - 1$ such that $ri + j \leq n$ we connect node $r \cdot i + j$ to node $r \cdot (i - 1) + j$.

3. For all $0 \leq i \neq j$ for which $ri + j, rj + i < n$ we connect nodes $r \cdot i + j$ and $r \cdot j + i$.

For all $0 \leq j \leq r-1$ we can contract all nodes $r \cdot i + j$ into a single node $\hat{j}$, by successively contracting edges (of type (ii)) $(r \cdot (i-1) + j, r \cdot i + j)$.

The contracted graph has $r$ nodes, and any two of its nodes $\hat{i} \neq \hat{j}$ are connected, because of the edge between nodes $r \cdot i + j$ and $r \cdot j + i$ in graph $D_r$. Thus the contracted graph is the complete graph $K_r$, thus $h(D_r) \geq r$.

One can verify that, because of the way graph $D_{r,n}$ is constructed $d(v) \leq 3$ for every $v \in V(G)$. Thus $\Delta(D_{r,n}) + 1 = 4$, $\Delta_2(D_{r,n}) + 1 = 4$ as well. As for the last bound, it is certainly no larger than $\Delta(D_{r,n}) + 2 = 5$.

Graph $D_{r,n}$ is 3-colorable, for any $r \geq 3$, because it is *2-degenerate*: in the natural ordering of the vertices each vertex is adjacent to at most two earlier nodes. Thus the GREEDY algorithm 3-colors $D_{r,n}$.

### 4.3 Proof of Theorem 3

To obtain the result we have to prove the inequality

$$m \geq n + 1 + \binom{h(G)}{2} - h(G) \qquad (6)$$

Indeed, in the decomposition of graphs corresponding to obtaining $K_{h(G)}$ at least one of the following situations holds:

1. At least one component corresponding to a vertex is *not* a tree (including single points among trees), or

2. There exist two components connected by at least two edges.

Indeed, suppose neither (1) nor (2) applies. Then by contracting each tree component we would get an acyclic contraction to a $K_{h(G)}$ (including, possibly, the odd cycle $C_3$). Thus m counts at least one extra edge not counted on the right hand side of equation (5), so equation (6) holds. Now an argument similar to that in the proof of Theorem 1 completes the proof.

### 4.4 Proof of Theorem 4

Apply the fact that Hadwiger's function is interpolating to graphs $D_{\lfloor \sqrt{n} \rfloor, n}$. This implies the inequality

$$sup_{\{G \in \mathcal{C}; d_1, \ldots, d_n \leq 3\}} \omega_n(d_1, d_2, \ldots, d_n) \geq h(D_{\lfloor \sqrt{n} \rfloor, n}) \geq \lfloor \sqrt{n} \rfloor.$$